\documentclass[%
 reprint,
showpacs,preprintnumbers,
 amsmath,amssymb,
prl,
floatfix
]{revtex4-1}

\usepackage{graphicx}
\usepackage{xcolor}
\usepackage{dcolumn}
\usepackage{bm}
\usepackage{hyperref}
\hypersetup{
    colorlinks=true,
    linkcolor=blue,
    citecolor=blue,     
    urlcolor=blue,
}
\usepackage{amsmath}
\usepackage{makecell}

\allowdisplaybreaks

\begin{document}

\title{Transient Electronic Phase Separation During Metal-Insulator Transitions}

\author{Yin Shi}
 \email{yxs187@psu.edu}
\author{Long-Qing Chen}%
 \email{lqc3@psu.edu}
\affiliation{%
Department of Materials Sciences and Engineering, Pennsylvania State University, University Park, PA 16802, USA
}%

%
%

\date{\today}

\begin{abstract}
From thermodynamic analysis we demonstrate that during metal-insulator transitions in pure matters, a nonequilibrium homogeneous state may be unstable against charge density modulations with certain wavelengths, and thus evolves to the equilibrium phase through transient electronic phase separation.
This phase instability occurs as two inequalities between the first and the second derivatives of the free energy with respect to the order parameter are fulfilled.
The dominant wavelength of the modulated phase is also derived.
The computer simulation further confirms the theoretical derivation.
Employing the pre-established phase-field model of VO$_2$, we show that this transient electronic phase separation may take place in VO$_2$ upon photoexcitation.


\end{abstract}

\maketitle


Phase separation widely exists in Nature.
For example, a homogeneous liquid separates into a liquid-vapor mixture when it is mechanically unstable (the isothermal compressibility becomes negative); a homogeneous binary solution decomposes into two immiscible parts as the chemical instability is reached (the system is inside the spinodal curve)~\cite{Landau80Statistical}.

In multinary complex materials, the interplay among charge/spin/lattice degrees of freedom may lead to competing ground states with distinct electronic properties~\cite{Cheong02Electronic,Moreo99Phase,Yunoki98Phase,Dagotto98Ferromagnetic}.
These states can coexist at low temperatures on microscopic length scale whereas any homogeneous phase is unstable, causing the electronic phase separation~\cite{Cheong02Electronic,Moreo99Phase,Yunoki98Phase,Dagotto98Ferromagnetic}, e.g., the phase separation into metal-insulator mixtures underlying the colossal magnetoresistance~\cite{Uehara99Percolative,Dagotto01Colossal}.
Like in the mechanical and the chemical phase separations, the electronic phase separation leads to stable phase mixtures.
In this work we report a \emph{transient} electronic phase separation into \emph{nonequilibrium} metal-insulator coexistence during metal-insulator transitions (MITs) in pure matters. 
Based on thermodynamic analysis taking into account the influence of free charges on the MIT, it is shown to result from the instability of a nonequilibrium homogeneous state against charge density modulations with certain wavelengths.

Vanadium dioxide ($\mathrm{VO}_2$) is a simple compound exhibiting MIT~\cite{Morin59Oxides,Zylbersztejn75Metal}, which may be a testbed for the transient electronic phase separation.
Below a temperature $T_c=338~\mathrm{K}$, $\mathrm{VO}_2$ is an insulator with a monoclinic structure (the M1 phase), while above $T_c$ it is a metal with a rutile structure (the R phase)~\cite{Morin59Oxides,Zylbersztejn75Metal,Park13Measurement}.
In previous works we have formulated a phase-field model of the MIT in $\mathrm{VO}_2$, in which the thermodynamics is described by a Landau potential as a functional of electronic and structural order parameter fields and free electron and hole density fields~\cite{Shi17Ginzburg,Shi18Phase,Shi19Current}.
Applying this model to the photoexcited $\mathrm{VO}_2$, we show that the transient metal-insulator coexistence may emerge from the nonequilibrium state induced by the photoexcitation.

In general a MIT can be characterized by an order parameter field, say, $\nu(\mathbf{r},t)$, which is a real number depending on the spatial coordinates $\mathbf{r}=(x,y,z)$ and the time $t$.
$\nu=0$ and $\nu\neq 0$ characterize the metal and the insulator, respectively (we assume that $\pm \nu$ correspond to different variants of the same phase).
The thermodynamics of the MIT is described by a (nonequilibrium) free energy functional consisting of a contribution from the intrinsic material $F_0$ and that from the additional free electrons and holes $F_f$,
\begin{align}
F[T;\nu(\mathbf{r},t),n(\mathbf{r},t),&p(\mathbf{r},t)] = F_0[T;\nu(\mathbf{r},t)] \notag  \\
&+F_f[T;\nu(\mathbf{r},t),n(\mathbf{r},t),p(\mathbf{r},t)],
\label{eq:F}
\end{align}
where $T$ is the temperature and $n(\mathbf{r},t)$ and $p(\mathbf{r},t)$ are the free electron and hole density fields, respectively.
$F_0$ is composed of a bulk energy term $f_b$ and a gradient energy term,
\begin{equation}
F_0=\int\left[f_b(T;\nu)+\frac{\kappa}{2}(\nabla\nu)^2\right]d^3r,
\label{eq:F0}
\end{equation}
where $\kappa$ is a positive constant.

$F_f$ can be constructed as follows.
For simplicity, we assume that the energy gap symmetrically opens with respect to the Fermi level of the metallic phase during the MIT (which is the case for VO$_2$~\cite{Miller12Unusually}) and approximate the electron and the hole densities using the Boltzmann statistics.
We shall follow the approximation adopted in the semiconductor physics that the conduction and the valence bands are effectively parabolic with effective densities of states $N_c$ and $N_v$, respectively~\cite{Moll64Physics}.
Hereafter the energy reference is chosen at the midpoint of the gap.
Then the electron and the hole densities are expressed as $n=N_c\exp[-(E_g/2-\mu_e)/k_BT]$ and $p=N_v\exp[-(E_g/2-\mu_h)/k_BT]$, respectively, where $E_g$ is the gap, $\mu_{e(h)}$ is the (quasi-) chemical potential of the electrons (holes), and $k_B$ is the Boltzmann constant~\cite{Moll64Physics}.
$F_f$ is thus calculated as
\begin{align}
F_f=& \int\left[ \int_0^n(\mu_e)_{TV}dn + \int_0^p(\mu_h)_{TV}dp \right]d^3r - F_i \notag   \\
      =& \int\bigg[ k_BT\left(n\ln\frac{n}{N_c}-n+p\ln\frac{p}{N_v}-p\right) \notag  \\
       &~~~~~~+\frac{E_g}{2}(n+p) \bigg]d^3r - F_i,
\label{eq:Ff}
\end{align}
where $V$ indicates the volume.
$F_i=-2\int k_BTn_id^3r$ is the equilibrium intrinsic free energy of the free electrons and holes, where $n_i$ is the equilibrium intrinsic carrier density.
Hence $F_f=0$ at the equilibrium intrinsic case.
Naturally the energy gap $E_g$ is related to the order parameter $\nu$.
Since $E_g$ is a scalar invariant with respect to the reverse of the sign of $\nu$, the symmetry-allowed expansion for $E_g$ on the lowest order of $\nu$ is $E_g=\gamma \nu^2$, where $\gamma$ is a positive constant.

The kinetics of the MIT is described by the Allen-Cahn equation for the non-conserved order parameter $\nu$, $\dot{\nu}=-L\delta F/\delta \nu$, and the diffusion equation for $n$, $\dot{n}=\nabla\cdot[(Mn/e)\nabla(\delta F/\delta n)]$~\cite{Chen02Phase}. Here we assume charge neutrality everywhere ($n=p$ always and thus no net electric field produced) and that the timescale of the transient electronic phase separation is much shorter than the lifetime of the free carriers (the source in the diffusion equation is ignored within the timescale considered).
The overhead dot represents the time derivative, $L$ is a positive constant related to the interface mobility, $M$ is the mobility of the electrons, and $e$ is the elementary charge.
We assume that the phase transition is much faster than the diffusion process of the electrons (that is, $\nu$ is in equilibrium at any moment for a given distribution of $n$).
In one dimension, these equations are then
\begin{align}
\frac{\partial f_b}{\partial \nu}-\kappa\nu''+2\gamma\nu (n-n_i)&=0,  \label{eq:cahn}  \\
\frac{M\gamma}{e}(\nu\nu' n'+n\nu'^2+n\nu\nu'')+\frac{Mk_BT}{e}n''&=\dot{n},   \label{eq:hilliard}
\end{align}
where the prime represents the spatial derivative along the $x$ dimension.

We now examine the stability of Eqs.~(\ref{eq:cahn}-\ref{eq:hilliard}) against infinitesimal fluctuations.
First, one can solve $n$ out as a function of $\nu$ from Eq.~(\ref{eq:cahn}), and substitute it into Eq.~(\ref{eq:hilliard}) to obtain a differentiation equation of $\nu$ only.
Then one may consider the solution to be a uniform value $\bar{\nu}$ plus an infinitesimal fluctuation with a wavenumber $k$,
\begin{equation}
\nu=\bar{\nu}+\psi_k(t)\exp(ikx),
\label{eq:sol}
\end{equation}
where $\psi_k$ is an infinitesimal amplitude.
This is also equivalent to an infinitesimal modulation of the electron (hole) density field.
Substitution of Eq.~(\ref{eq:sol}) into Eqs.~(\ref{eq:cahn}-\ref{eq:hilliard}) gives, to the first order of $\psi_k$, 
\begin{equation}
\dot{\psi}_k=R(k)\psi_k,
\label{eq:psi}
\end{equation}
with
\begin{equation}
R(k)=-\frac{M}{e}\frac{h_1k^2+h_2k^4}{h_3+h_4k^2},
\label{eq:R}
\end{equation}
where
\begin{subequations}
\begin{align}
h_1=&\frac{k_BT}{\gamma}\frac{\partial^2f_b}{\partial\nu^2}\bigg|_{\bar{\nu}}+\left( \bar{\nu}-\frac{k_BT}{\gamma\bar{\nu}} \right)\frac{\partial f_b}{\partial \nu}\bigg|_{\bar{\nu}},  \\
h_2=&\frac{k_BT\kappa}{\gamma},  \\
h_3=&\frac{2\gamma\bar{\nu}^2\bar{n}_i}{k_BT}+\frac{1}{\gamma}\left( \frac{\partial^2f_b}{\partial\nu^2}\bigg|_{\bar{\nu}}-\frac{1}{\bar{\nu}}\frac{\partial f_b}{\partial \nu}\bigg|_{\bar{\nu}} \right),  \\
h_4=&\frac{\kappa}{\gamma}.
\end{align}
\end{subequations}
$\bar{n}_i$ is $n_i$ at $\bar{\nu}$.
The solution to Eq.~(\ref{eq:psi}) is just $\psi_k(t)=\exp[R(k)t]$.
Hence $R<0$ indicates that $\psi_k$ vanishes with time and Eqs.~(\ref{eq:cahn}-\ref{eq:hilliard}) are stable against small fluctuations, while $R>0$ indicates that $\psi_k$ grows exponentially with the time and Eqs.~(\ref{eq:cahn}-\ref{eq:hilliard}) are unstable against small fluctuations.

\begin{figure*}[t]
 \includegraphics[width=0.98\textwidth]{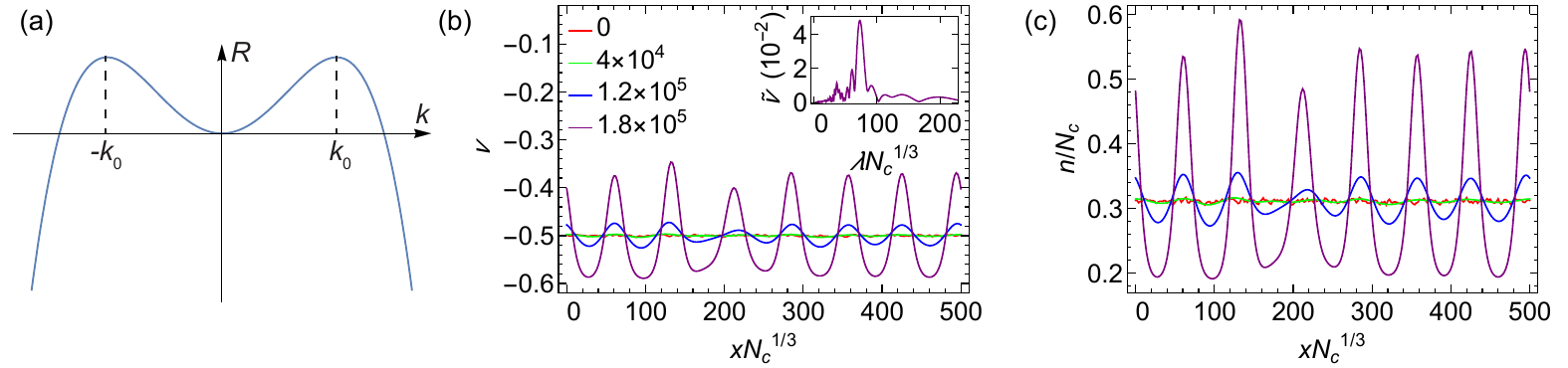}
 \caption{\label{fig:theory} Transient electronic phase separation during a MIT. (a) Shape of $R(k)$ inside the unstable regime. (b) and (c) are the temporal evolution of $\nu$ and $n$, respectively. The legend in (b) lists the times in units of $e/Mk_BT_0N_c^{2/3}$. The inset in (b) is the Fourier transformation of $\nu$ at $t=1.8\times 10^5e/Mk_BT_0N_c^{2/3}$ ($\lambda$ is the wavelength).}
\end{figure*}

For small $k$, we can expand Eq.~(\ref{eq:R}) to the fourth order of $k$,
\begin{equation}
R(k)\approx -\frac{M}{e}\left( \frac{h_1}{h_3}k^2+\frac{h_2h_3-h_1h_4}{h_3^2}k^4 \right).
\label{eq:R4}
\end{equation}
If $h_1/h_3>0$ and $h_2h_3-h_1h_4>0$, $R(k)$ in Eq.~(\ref{eq:R4}) is always nonpositive, which corresponds to the stable regime.
One can check that the equilibrium case [$(\partial f_b/\partial \nu)|_{\bar{\nu}}=0,(\partial^2 f_b/\partial \nu^2)|_{\bar{\nu}}>0$] is included in the stable regime, as expected.
On the other hand, if $h_1/h_3<0$ and $h_2h_3-h_1h_4>0$, or equivalently,
\begin{equation}
\frac{1}{\bar{\nu}}\frac{\partial f_b}{\partial \nu}\bigg|_{\bar{\nu}}-\frac{2\gamma^2\bar{\nu}^2\bar{n}_i}{k_BT}<\frac{\partial^2 f_b}{\partial \nu^2}\bigg|_{\bar{\nu}}<\left( 1-\frac{\gamma\bar{\nu}^2}{k_BT} \right)\frac{1}{\bar{\nu}}\frac{\partial f_b}{\partial \nu}\bigg|_{\bar{\nu}},
\label{eq:unstable}
\end{equation}
$R(k)$ in Eq.~(\ref{eq:R4}) has a shape as shown in Fig.~\ref{fig:theory}(a). 
As can be seen, the unstable range of $k$, i.e., those $k$ yielding $R(k)>0$, is
\begin{equation}
|k|<\sqrt{\frac{-h_1h_3}{h_2h_3-h_1h_4}}, k\neq 0.
\end{equation}
Inequalities~(\ref{eq:unstable}) thus correspond to the unstable regime.
$R(k)$ has two positive maxima at $\pm k_0$, with $k_0$ calculated to be $\sqrt{-h_1h_3/2(h_2h_3-h_1h_4)}$.
Since $\psi_k$ grows exponentially with the time at the rate $R(k)$, the dominant $\psi_k$'s are those at $k=\pm k_0$, which leads to a modulation of $\nu$ with the wavelength 
\begin{equation}
\lambda_0=\frac{2\pi}{k_0}=2\pi\sqrt{\frac{2(h_2h_3-h_1h_4)}{-h_1h_3}}.
\label{eq:wavelength}
\end{equation}
Therefore, for a $\bar{\nu}$ satisfying condition~(\ref{eq:unstable}), the initial homogeneous phase with the order parameter $\bar{\nu}$ will spontaneously separate to the mixture of a metal-like phase and an insulator-like phase.

To confirm the above derivation, we numerically solve Eqs.~(\ref{eq:cahn}-\ref{eq:hilliard}) with periodic boundary conditions for both $\nu$ and $n$.
We use the standard Landau polynomial for $f_b$, $f_b=-4f_0(\tau\nu^2/2+\nu^4/4)$ with $\tau=(T-T_0)/T_0$, which describes a second-order phase transition at a critical temperature $T_0$.
$f_0$ is the equilibrium free energy density at $T=0$~K.
The parameters are chosen to be $f_0=-0.25k_BT_0N_c,\kappa=8k_BT_0N_c^{1/3}$ and $\gamma=5k_BT_0$.

The calculation result at temperature $\tau=-0.5$ is shown in Fig.~\ref{fig:theory}(b-c).
Initially $\nu$ has a value $\bar{\nu}=-0.5$ plus a random noise ranging from $-0.005$ to $0.005$.
This $\bar{\nu}$ is inside the unstable regime (\ref{eq:unstable}).
Indeed, the noise grows with the time with a dominant wavelength, which is shown clearly in the Fourier transformation of $\nu$ ($\tilde{\nu}$) at $t=1.8\times 10^5 e/Mk_BT_0N_c^{2/3}$.
The highest peak in $\tilde{\nu}$ is at the wavelength $\lambda_0=74N_c^{-1/3}$, which is very close to the $\lambda_0=72N_c^{-1/3}$ calculated from Eq.~(\ref{eq:wavelength}).
We also observe the coarsening of the metal-like phase (peaks of $\nu$) and the insulator-like phase (valleys of $\nu$) at later stages (not shown).
If adding to Eq.~(\ref{eq:hilliard}) the source term representing the electron-hole recombination process (the process for $n$ to approach $n_i$), we see that the transient metal-insulator mixture eventually evolves to the equilibrium homogeneous insulator with $\nu=-\sqrt{-\tau}=-1/\sqrt{2}$.
The simulation of the cases inside the stable regime shows that the initial noise shrinks and that the system directly evolves to the equilibrium homogeneous insulator.

In the spinodal decomposition in binary solutions, the solute-concentration modulation with a finite wavelength rises from the interaction of the concentration (the gradient energy)~\cite{Cahn61On}.
In the transient electronic phase separation, however, the free energy of free electrons and holes in Eq.~(\ref{eq:Ff}) does not possess the gradient energy; rather, the coupling of Eq.~(\ref{eq:cahn}) and Eq.~(\ref{eq:hilliard}) gives rise to an effective interaction of the free-electron and free-hole densities.

In the above derivation, it is seen that the electronic phase separation occurs only when the initial state is nonequilibrium.
We expect that the ultrafast photoexcitation may possibly lead to this electronic phase separation, since it can drive a system away from equilibrium.
As an example we will examine the photoinduced MIT in $\mathrm{VO}_2$.

\begin{figure*}[t]
 \includegraphics[width=0.98\textwidth]{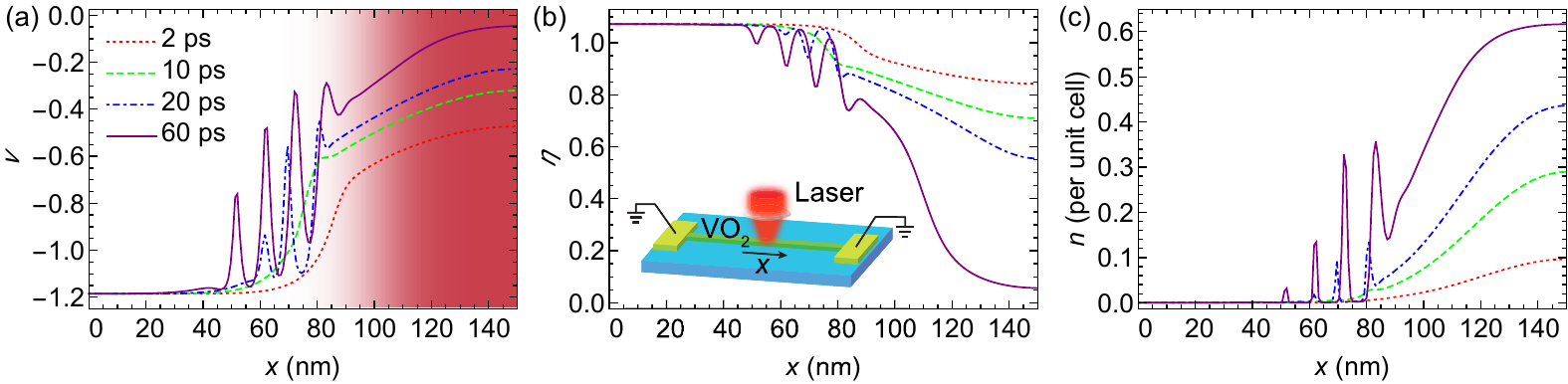}
 \caption{\label{fig:VO2} Transient electronic phase separation in a $300$-nm-long $\mathrm{VO}_2$ nanobeam at $T=320~\mathrm{K}$ photoexcited by an $800$-nm laser pulse ($I_0=10^7~\mathrm{W/cm^2},~\delta=25~\mathrm{nm},~x_0=150~\mathrm{nm},~\zeta=0.1~\mathrm{ns},~t_0=0.2~\mathrm{ns}$). (a), (b) and (c) are the temporal evolution of $\nu$, $\eta$ and $n$, respectively. $p$ is almost the same as $n$. The inset in (b) is the schematic of the simulation setup. We only show the results on half of the sample ($0\leq x\leq 150~\mathrm{nm}$) [the shaded region in (a) indicates the half of the illuminated region]. The results on the other half is symmetrical with the shown results about the $x=150~\mathrm{nm}$ mirror plane.}
\end{figure*}

The MIT in $\mathrm{VO}_2$ can be described by two order parameter fields $\eta(\mathbf{r},t)$ and $\nu(\mathbf{r},t)$, which characterize the structural and the electronic phases, respectively~\cite{Shi17Ginzburg,Shi18Phase,Shi19Current,Sup}.
The R and the M1 phases correspond to $\nu=0,\eta=0$ and $\nu\neq 0,\eta\neq 0,\nu\eta<0$, respectively.
Like in obtaining Eqs.~(\ref{eq:cahn}-\ref{eq:hilliard}), the kinetics of the phase transitions in $\mathrm{VO}_2$ under photoexcitation is governed by the Allen-Cahn equations for $\eta$ and $\nu$ and the diffusion equations for $n$ and $p$,
\begin{align}
\dot{\eta}(\mathbf{r},t)=& -L_1\frac{\delta F_{\mathrm{VO}_2}}{\delta \eta(\mathbf{r},t)},  \label{eq:etaev}  \\
\dot{\nu}(\mathbf{r},t)=& -L_2\frac{\delta F_{\mathrm{VO}_2}}{\delta \nu(\mathbf{r},t)},  \label{eq:nuev}  \\
\dot{n}(\mathbf{r},t)=&\nabla\cdot\left[ \frac{M_en(\mathbf{r},t)}{e}\nabla\frac{\delta F_{\mathrm{VO}_2}}{\delta n(\mathbf{r},t)} \right]+\Gamma(\mathbf{r},t),  \label{eq:nev}  \\
\dot{p}(\mathbf{r},t)=&\nabla\cdot\left[ \frac{M_hp(\mathbf{r},t)}{e}\nabla\frac{\delta F_{\mathrm{VO}_2}}{\delta p(\mathbf{r},t)} \right]+\Gamma(\mathbf{r},t),  \label{eq:pev}
\end{align}
which are closed by the Poisson equation for the self-consistent determination of the electric potential $\Phi(\mathbf{r},t)$, 
\begin{equation}
-\nabla^2 \Phi(\mathbf{r},t)=\frac{e[p(\mathbf{r},t)-n(\mathbf{r},t)]}{\epsilon_r\epsilon_0}.
\label{eq:Poisson}
\end{equation}
Here $F_{\mathrm{VO}_2}=F_{\mathrm{VO}_2}[T,\Phi(\mathbf{r},t);\eta(\mathbf{r},t),\nu(\mathbf{r},t),n(\mathbf{r},t),p(\mathbf{r},t)]$ is the Landau free energy of $\mathrm{VO}_2$~\cite{Sup}.
$\Gamma(\mathbf{r},t)$ represents the photoexcitation rate of free electron-hole pairs.
$L_1$ and $L_2$ are constants related to the interface mobility, and $M_{e(h)}$ is the electron (hole)  mobility.
$\epsilon_r$ and $\epsilon_0$ are the relative permittivity of $\mathrm{VO}_2$ and the vacuum permittivity, respectively.
The electron-hole recombination process can be ignored here since the lifetime of free electrons and holes in $\mathrm{VO}_2$ ($\sim 10~\mathrm{\mu s}$~\cite{Miller12Unusually}) is found to be much longer than the timescale of the transient electronic phase separation.
For a monochome light with an angular frequency $\omega$ and an intensity $I$, $\Gamma$ can be derived from the Fermi's golden rule~\cite{Sup},
\begin{align}
\Gamma=& \frac{\sqrt{2\pi N_vN_c}e^2E_gI}{\epsilon_0cm\omega^2\hbar k_BT}\left( 1+\frac{m}{m_h^*} \right)\sqrt{\frac{\hbar\omega-E_g}{k_BT}}  \notag   \\
& \times f\left(-\frac{\hbar\omega}{2}+\mu_h\right)\left[1-f\left(\frac{\hbar\omega}{2}-\mu_e\right) \right],
\end{align}
where $m_h^*$ is the effective mass of holes in $\mathrm{VO}_2$, $c$ is the speed of light in the vacuum, $m$ is the electron mass, $\hbar$ is the Planck constant divided by $2\pi$, and $f(\varepsilon)=[1+\exp(\varepsilon/k_BT)]^{-1}$ is the Fermi distribution function.
$I$ is a Gaussian-type function of both the space and the time controlling the illumination range and duration of the pump laser pulse: in the 1D case $I(x,t)=\sqrt{\mathrm{e}}I_0 g_\delta(x-x_0) g_\zeta(t-t_0)$, where $g_\sigma(\epsilon)=\exp(-\epsilon^2/2\sigma^2)$ and $I_0$ is defined as the intensity of the laser pulse.
The illumination width and the pulse duration are defined as $4\delta$ and $2\zeta$, respectively.
$x_0$ and $t_0$ are the position and the moment of the peak of the laser pulse, respectively.

\begin{figure}[b]
 \includegraphics[width=0.33\textwidth]{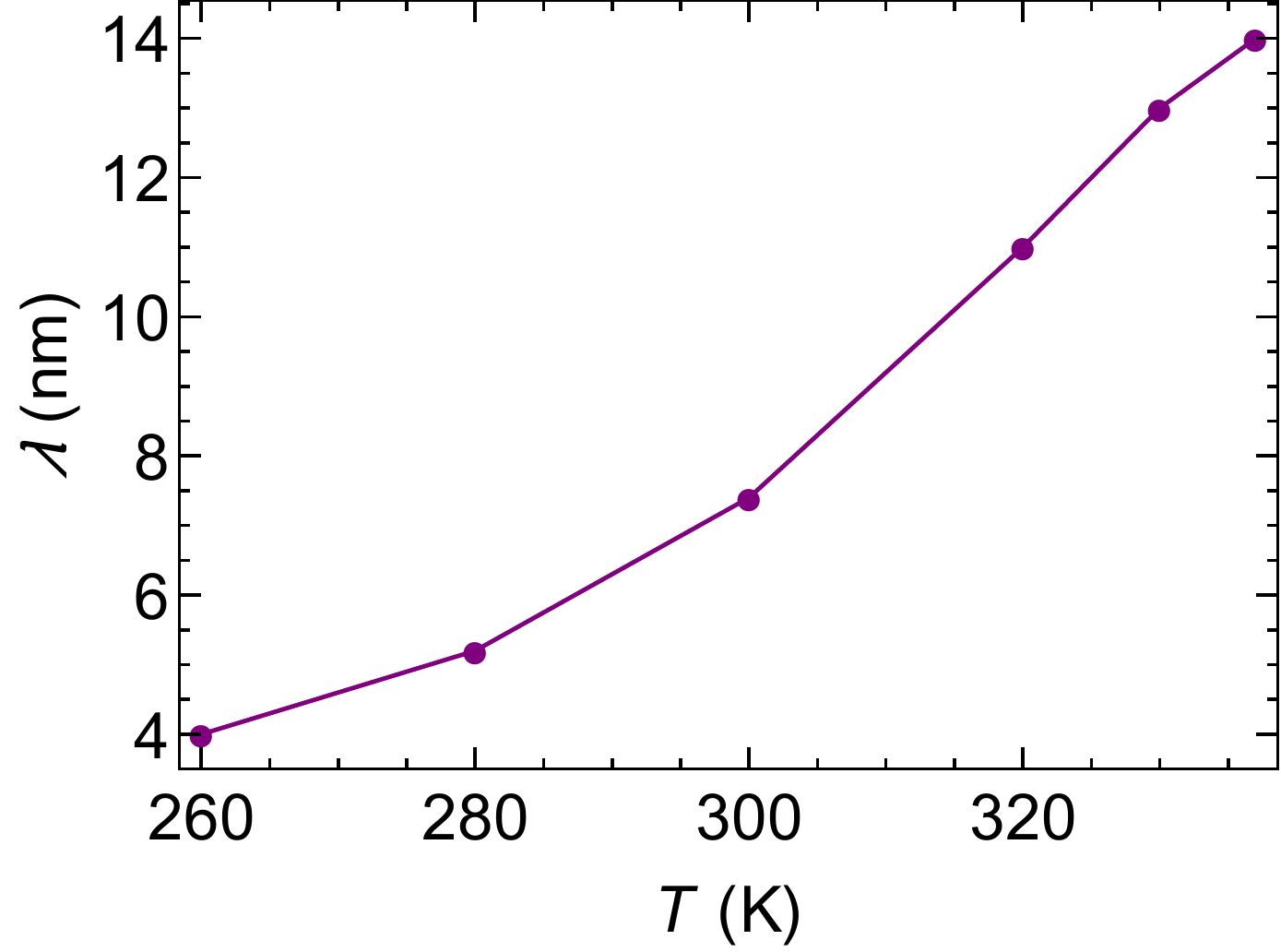}
 \caption{\label{fig:wavelen} Wavelength of the phase modulation as a function of the temperature in the photoexcitated $\mathrm{VO}_2$. The fixed conditions are the same as in Fig.~\ref{fig:VO2}. The line is guide to eyes.}
\end{figure}

The simulation setup is sketched in the inset in Fig. \ref{fig:VO2}(b).
We consider a $\mathrm{VO}_2$ nanobeam (1D system) with its two ends connected to the ground.
The corresponding boundary conditions for Eqs.~(\ref{eq:etaev}-\ref{eq:Poisson}) are that $\Phi$ is zero and $n,p$ have their equilibrium values at the two ends.
We assume Neumann boundary conditions (zero spatial derivatives) for $\eta$ and $\nu$ at that two ends.
Figure~\ref{fig:VO2} presents the calculation results for the photoexcited $\mathrm{VO}_2$.
Inside the illuminated region, the photoexcitation produces free electron-hole pairs, which screen the electron-electron repulsion.
This eventually leads to the closure of the gap, resulting in the transition from the insulator to the metal inside the illuminated region.
On the other hand, the free electrons and holes diffuse outside the illuminated region, inducing a nonequilibrium state there.
As can be seen clearly, a phase and a charge density modulations with a wavelength of $\sim 11~\mathrm{nm}$ take place around $x\sim 70~\mathrm{nm}$ several picoseconds after the incidence of the laser, inducing coexistence of the metal-like (high carrier density) and the insulator-like (low carrier density) phases.
A wave-like ionic displacement field (represented by $\eta$) also develops around $x\sim 70~\mathrm{nm}$ due to the coupling between the electronic and the structural order parameters.
The metal-like and the insulator-like phases coarsen at later stages.

We further calculate the wavelength of the phase modulation at various temperatures, as shown in Fig.~\ref{fig:wavelen}.
The wavelength increases with elevating temperature, and reaches the maximum of $\sim 14~\mathrm{nm}$ near $T_c$.
At low temperatures the modulation wavelength may be calculated to be sub-nanometers.
This should be considered as invalid since the Landau theory is essentially not applicable to the lengthscale comparable to the lattice constant.

In summary, we have shown theoretically that in materials exhibiting MIT, a homogeneous state away from equilibrium may be unstable against charge density modulations with a certain range of wavelengths.
Hence the homogeneous state evolves to the equilibrium phase not homogeneously, but through transient electronic phase separation with a dominant wavelength.
The criterion for the onset of this phase instability has something to do with both the first and the second derivatives of the free energy with respect to the order parameter.

Employing the phase-field model of $\mathrm{VO}_2$ that has been extended to the photoexcitation problem, we have shown that the transient electronic phase separation may emerge in photoexcited $\mathrm{VO}_2$ at the edge of the illuminated region, which is accompanied by a transient wave-like ionic displacement field at the same place.
Further experiments may pay attention to detecting the transient electronic phase separation and the modulated charge density and ionic displacement fields in $\mathrm{VO}_2$ and other materials exhibiting MIT.

This work was funded by the Penn State MRSEC, Center for Nanoscale Science, under the award NSF DMR-1420620 and DMR-1744213.

\bibliography{VO2spidec_ref}

\begin{thebibliography}{18}%
\makeatletter
\providecommand \@ifxundefined [1]{%
 \@ifx{#1\undefined}
}%
\providecommand \@ifnum [1]{%
 \ifnum #1\expandafter \@firstoftwo
 \else \expandafter \@secondoftwo
 \fi
}%
\providecommand \@ifx [1]{%
 \ifx #1\expandafter \@firstoftwo
 \else \expandafter \@secondoftwo
 \fi
}%
\providecommand \natexlab [1]{#1}%
\providecommand \enquote  [1]{``#1''}%
\providecommand \bibnamefont  [1]{#1}%
\providecommand \bibfnamefont [1]{#1}%
\providecommand \citenamefont [1]{#1}%
\providecommand \href@noop [0]{\@secondoftwo}%
\providecommand \href [0]{\begingroup \@sanitize@url \@href}%
\providecommand \@href[1]{\@@startlink{#1}\@@href}%
\providecommand \@@href[1]{\endgroup#1\@@endlink}%
\providecommand \@sanitize@url [0]{\catcode `\\12\catcode `\$12\catcode
  `\&12\catcode `\#12\catcode `\^12\catcode `\_12\catcode `\%12\relax}%
\providecommand \@@startlink[1]{}%
\providecommand \@@endlink[0]{}%
\providecommand \url  [0]{\begingroup\@sanitize@url \@url }%
\providecommand \@url [1]{\endgroup\@href {#1}{\urlprefix }}%
\providecommand \urlprefix  [0]{URL }%
\providecommand \Eprint [0]{\href }%
\providecommand \doibase [0]{http://dx.doi.org/}%
\providecommand \selectlanguage [0]{\@gobble}%
\providecommand \bibinfo  [0]{\@secondoftwo}%
\providecommand \bibfield  [0]{\@secondoftwo}%
\providecommand \translation [1]{[#1]}%
\providecommand \BibitemOpen [0]{}%
\providecommand \bibitemStop [0]{}%
\providecommand \bibitemNoStop [0]{.\EOS\space}%
\providecommand \EOS [0]{\spacefactor3000\relax}%
\providecommand \BibitemShut  [1]{\csname bibitem#1\endcsname}%
\let\auto@bib@innerbib\@empty
\bibitem [{\citenamefont {Landau}\ and\ \citenamefont
  {Lifshitz}(1980)}]{Landau80Statistical}%
  \BibitemOpen
  \bibfield  {author} {\bibinfo {author} {\bibfnamefont {L.}~\bibnamefont
  {Landau}}\ and\ \bibinfo {author} {\bibfnamefont {E.}~\bibnamefont
  {Lifshitz}},\ }\href@noop {} {\emph {\bibinfo {title} {Statistical Physics,
  Part 1: Volume 5}}},\ \bibinfo {edition} {3rd}\ ed.\ (\bibinfo  {publisher}
  {Butterworth-Heinemann},\ \bibinfo {address} {Oxford},\ \bibinfo {year}
  {1980})\BibitemShut {NoStop}%
\bibitem [{\citenamefont {Cheong}\ \emph {et~al.}(2002)\citenamefont {Cheong},
  \citenamefont {Sharma}, \citenamefont {Hur}, \citenamefont {Horibe},\ and\
  \citenamefont {Chen}}]{Cheong02Electronic}%
  \BibitemOpen
  \bibfield  {author} {\bibinfo {author} {\bibfnamefont {S.-W.}\ \bibnamefont
  {Cheong}}, \bibinfo {author} {\bibfnamefont {P.}~\bibnamefont {Sharma}},
  \bibinfo {author} {\bibfnamefont {N.}~\bibnamefont {Hur}}, \bibinfo {author}
  {\bibfnamefont {Y.}~\bibnamefont {Horibe}}, \ and\ \bibinfo {author}
  {\bibfnamefont {C.}~\bibnamefont {Chen}},\ }\href {\doibase
  https://doi.org/10.1016/S0921-4526(02)00772-X} {\bibfield  {journal}
  {\bibinfo  {journal} {Physica B: Condensed Matter}\ }\textbf {\bibinfo
  {volume} {318}},\ \bibinfo {pages} {39 } (\bibinfo {year}
  {2002})}\BibitemShut {NoStop}%
\bibitem [{\citenamefont {Moreo}\ \emph {et~al.}(1999)\citenamefont {Moreo},
  \citenamefont {Yunoki},\ and\ \citenamefont {Dagotto}}]{Moreo99Phase}%
  \BibitemOpen
  \bibfield  {author} {\bibinfo {author} {\bibfnamefont {A.}~\bibnamefont
  {Moreo}}, \bibinfo {author} {\bibfnamefont {S.}~\bibnamefont {Yunoki}}, \
  and\ \bibinfo {author} {\bibfnamefont {E.}~\bibnamefont {Dagotto}},\ }\href
  {\doibase 10.1126/science.283.5410.2034} {\bibfield  {journal} {\bibinfo
  {journal} {Science}\ }\textbf {\bibinfo {volume} {283}},\ \bibinfo {pages}
  {2034} (\bibinfo {year} {1999})}\BibitemShut {NoStop}%
\bibitem [{\citenamefont {Yunoki}\ \emph {et~al.}(1998)\citenamefont {Yunoki},
  \citenamefont {Hu}, \citenamefont {Malvezzi}, \citenamefont {Moreo},
  \citenamefont {Furukawa},\ and\ \citenamefont {Dagotto}}]{Yunoki98Phase}%
  \BibitemOpen
  \bibfield  {author} {\bibinfo {author} {\bibfnamefont {S.}~\bibnamefont
  {Yunoki}}, \bibinfo {author} {\bibfnamefont {J.}~\bibnamefont {Hu}}, \bibinfo
  {author} {\bibfnamefont {A.~L.}\ \bibnamefont {Malvezzi}}, \bibinfo {author}
  {\bibfnamefont {A.}~\bibnamefont {Moreo}}, \bibinfo {author} {\bibfnamefont
  {N.}~\bibnamefont {Furukawa}}, \ and\ \bibinfo {author} {\bibfnamefont
  {E.}~\bibnamefont {Dagotto}},\ }\href {\doibase 10.1103/PhysRevLett.80.845}
  {\bibfield  {journal} {\bibinfo  {journal} {Phys. Rev. Lett.}\ }\textbf
  {\bibinfo {volume} {80}},\ \bibinfo {pages} {845} (\bibinfo {year}
  {1998})}\BibitemShut {NoStop}%
\bibitem [{\citenamefont {Dagotto}\ \emph {et~al.}(1998)\citenamefont
  {Dagotto}, \citenamefont {Yunoki}, \citenamefont {Malvezzi}, \citenamefont
  {Moreo}, \citenamefont {Hu}, \citenamefont {Capponi}, \citenamefont
  {Poilblanc},\ and\ \citenamefont {Furukawa}}]{Dagotto98Ferromagnetic}%
  \BibitemOpen
  \bibfield  {author} {\bibinfo {author} {\bibfnamefont {E.}~\bibnamefont
  {Dagotto}}, \bibinfo {author} {\bibfnamefont {S.}~\bibnamefont {Yunoki}},
  \bibinfo {author} {\bibfnamefont {A.~L.}\ \bibnamefont {Malvezzi}}, \bibinfo
  {author} {\bibfnamefont {A.}~\bibnamefont {Moreo}}, \bibinfo {author}
  {\bibfnamefont {J.}~\bibnamefont {Hu}}, \bibinfo {author} {\bibfnamefont
  {S.}~\bibnamefont {Capponi}}, \bibinfo {author} {\bibfnamefont
  {D.}~\bibnamefont {Poilblanc}}, \ and\ \bibinfo {author} {\bibfnamefont
  {N.}~\bibnamefont {Furukawa}},\ }\href {\doibase 10.1103/PhysRevB.58.6414}
  {\bibfield  {journal} {\bibinfo  {journal} {Phys. Rev. B}\ }\textbf {\bibinfo
  {volume} {58}},\ \bibinfo {pages} {6414} (\bibinfo {year}
  {1998})}\BibitemShut {NoStop}%
\bibitem [{\citenamefont {Uehara}\ \emph {et~al.}(1999)\citenamefont {Uehara},
  \citenamefont {Mori}, \citenamefont {Chen},\ and\ \citenamefont
  {Cheong}}]{Uehara99Percolative}%
  \BibitemOpen
  \bibfield  {author} {\bibinfo {author} {\bibfnamefont {M.}~\bibnamefont
  {Uehara}}, \bibinfo {author} {\bibfnamefont {S.}~\bibnamefont {Mori}},
  \bibinfo {author} {\bibfnamefont {C.}~\bibnamefont {Chen}}, \ and\ \bibinfo
  {author} {\bibfnamefont {S.-W.}\ \bibnamefont {Cheong}},\ }\href
  {https://www.nature.com/articles/21142} {\bibfield  {journal} {\bibinfo
  {journal} {Nature}\ }\textbf {\bibinfo {volume} {399}},\ \bibinfo {pages}
  {560} (\bibinfo {year} {1999})}\BibitemShut {NoStop}%
\bibitem [{\citenamefont {Dagotto}\ \emph {et~al.}(2001)\citenamefont
  {Dagotto}, \citenamefont {Hotta},\ and\ \citenamefont
  {Moreo}}]{Dagotto01Colossal}%
  \BibitemOpen
  \bibfield  {author} {\bibinfo {author} {\bibfnamefont {E.}~\bibnamefont
  {Dagotto}}, \bibinfo {author} {\bibfnamefont {T.}~\bibnamefont {Hotta}}, \
  and\ \bibinfo {author} {\bibfnamefont {A.}~\bibnamefont {Moreo}},\ }\href
  {\doibase https://doi.org/10.1016/S0370-1573(00)00121-6} {\bibfield
  {journal} {\bibinfo  {journal} {Physics Reports}\ }\textbf {\bibinfo {volume}
  {344}},\ \bibinfo {pages} {1 } (\bibinfo {year} {2001})}\BibitemShut
  {NoStop}%
\bibitem [{\citenamefont {Morin}(1959)}]{Morin59Oxides}%
  \BibitemOpen
  \bibfield  {author} {\bibinfo {author} {\bibfnamefont {F.~J.}\ \bibnamefont
  {Morin}},\ }\href {\doibase 10.1103/PhysRevLett.3.34} {\bibfield  {journal}
  {\bibinfo  {journal} {Phys. Rev. Lett.}\ }\textbf {\bibinfo {volume} {3}},\
  \bibinfo {pages} {34} (\bibinfo {year} {1959})}\BibitemShut {NoStop}%
\bibitem [{\citenamefont {Zylbersztejn}\ and\ \citenamefont
  {Mott}(1975)}]{Zylbersztejn75Metal}%
  \BibitemOpen
  \bibfield  {author} {\bibinfo {author} {\bibfnamefont {A.}~\bibnamefont
  {Zylbersztejn}}\ and\ \bibinfo {author} {\bibfnamefont {N.~F.}\ \bibnamefont
  {Mott}},\ }\href {\doibase 10.1103/PhysRevB.11.4383} {\bibfield  {journal}
  {\bibinfo  {journal} {Phys. Rev. B}\ }\textbf {\bibinfo {volume} {11}},\
  \bibinfo {pages} {4383} (\bibinfo {year} {1975})}\BibitemShut {NoStop}%
\bibitem [{\citenamefont {Park}\ \emph {et~al.}(2013)\citenamefont {Park},
  \citenamefont {Coy}, \citenamefont {Kasirga}, \citenamefont {Huang},
  \citenamefont {Fei}, \citenamefont {Hunter},\ and\ \citenamefont
  {Cobden}}]{Park13Measurement}%
  \BibitemOpen
  \bibfield  {author} {\bibinfo {author} {\bibfnamefont {J.~H.}\ \bibnamefont
  {Park}}, \bibinfo {author} {\bibfnamefont {J.~M.}\ \bibnamefont {Coy}},
  \bibinfo {author} {\bibfnamefont {T.~S.}\ \bibnamefont {Kasirga}}, \bibinfo
  {author} {\bibfnamefont {C.}~\bibnamefont {Huang}}, \bibinfo {author}
  {\bibfnamefont {Z.}~\bibnamefont {Fei}}, \bibinfo {author} {\bibfnamefont
  {S.}~\bibnamefont {Hunter}}, \ and\ \bibinfo {author} {\bibfnamefont {D.~H.}\
  \bibnamefont {Cobden}},\ }\href {https://www.nature.com/articles/nature12425}
  {\bibfield  {journal} {\bibinfo  {journal} {Nature}\ }\textbf {\bibinfo
  {volume} {500}},\ \bibinfo {pages} {431} (\bibinfo {year}
  {2013})}\BibitemShut {NoStop}%
\bibitem [{\citenamefont {Shi}\ \emph {et~al.}(2017)\citenamefont {Shi},
  \citenamefont {Xue},\ and\ \citenamefont {Chen}}]{Shi17Ginzburg}%
  \BibitemOpen
  \bibfield  {author} {\bibinfo {author} {\bibfnamefont {Y.}~\bibnamefont
  {Shi}}, \bibinfo {author} {\bibfnamefont {F.}~\bibnamefont {Xue}}, \ and\
  \bibinfo {author} {\bibfnamefont {L.-Q.}\ \bibnamefont {Chen}},\ }\href
  {http://stacks.iop.org/0295-5075/120/i=4/a=46003} {\bibfield  {journal}
  {\bibinfo  {journal} {Europhysics Letters}\ }\textbf {\bibinfo {volume}
  {120}},\ \bibinfo {pages} {46003} (\bibinfo {year} {2017})}\BibitemShut
  {NoStop}%
\bibitem [{\citenamefont {Shi}\ and\ \citenamefont {Chen}(2018)}]{Shi18Phase}%
  \BibitemOpen
  \bibfield  {author} {\bibinfo {author} {\bibfnamefont {Y.}~\bibnamefont
  {Shi}}\ and\ \bibinfo {author} {\bibfnamefont {L.-Q.}\ \bibnamefont {Chen}},\
  }\href {\doibase 10.1103/PhysRevMaterials.2.053803} {\bibfield  {journal}
  {\bibinfo  {journal} {Phys. Rev. Materials}\ }\textbf {\bibinfo {volume}
  {2}},\ \bibinfo {pages} {053803} (\bibinfo {year} {2018})}\BibitemShut
  {NoStop}%
\bibitem [{\citenamefont {Shi}\ and\ \citenamefont
  {Chen}(2019)}]{Shi19Current}%
  \BibitemOpen
  \bibfield  {author} {\bibinfo {author} {\bibfnamefont {Y.}~\bibnamefont
  {Shi}}\ and\ \bibinfo {author} {\bibfnamefont {L.-Q.}\ \bibnamefont {Chen}},\
  }\href {\doibase 10.1103/PhysRevApplied.11.014059} {\bibfield  {journal}
  {\bibinfo  {journal} {Phys. Rev. Applied}\ }\textbf {\bibinfo {volume}
  {11}},\ \bibinfo {pages} {014059} (\bibinfo {year} {2019})}\BibitemShut
  {NoStop}%
\bibitem [{\citenamefont {Miller}\ \emph {et~al.}(2012)\citenamefont {Miller},
  \citenamefont {Triplett}, \citenamefont {Lammatao}, \citenamefont {Suh},
  \citenamefont {Fu}, \citenamefont {Wu},\ and\ \citenamefont
  {Yu}}]{Miller12Unusually}%
  \BibitemOpen
  \bibfield  {author} {\bibinfo {author} {\bibfnamefont {C.}~\bibnamefont
  {Miller}}, \bibinfo {author} {\bibfnamefont {M.}~\bibnamefont {Triplett}},
  \bibinfo {author} {\bibfnamefont {J.}~\bibnamefont {Lammatao}}, \bibinfo
  {author} {\bibfnamefont {J.}~\bibnamefont {Suh}}, \bibinfo {author}
  {\bibfnamefont {D.}~\bibnamefont {Fu}}, \bibinfo {author} {\bibfnamefont
  {J.}~\bibnamefont {Wu}}, \ and\ \bibinfo {author} {\bibfnamefont
  {D.}~\bibnamefont {Yu}},\ }\href {\doibase 10.1103/PhysRevB.85.085111}
  {\bibfield  {journal} {\bibinfo  {journal} {Phys. Rev. B}\ }\textbf {\bibinfo
  {volume} {85}},\ \bibinfo {pages} {085111} (\bibinfo {year}
  {2012})}\BibitemShut {NoStop}%
\bibitem [{\citenamefont {Moll}(1964)}]{Moll64Physics}%
  \BibitemOpen
  \bibfield  {author} {\bibinfo {author} {\bibfnamefont {J.~L.}\ \bibnamefont
  {Moll}},\ }\href@noop {} {\emph {\bibinfo {title} {{Physics of
  semiconductors}}}}\ (\bibinfo  {publisher} {McGraw-Hill},\ \bibinfo {address}
  {New York},\ \bibinfo {year} {1964})\BibitemShut {NoStop}%
\bibitem [{\citenamefont {Chen}(2002)}]{Chen02Phase}%
  \BibitemOpen
  \bibfield  {author} {\bibinfo {author} {\bibfnamefont {L.-Q.}\ \bibnamefont
  {Chen}},\ }\href
  {https://www.annualreviews.org/doi/10.1146/annurev.matsci.32.112001.132041}
  {\bibfield  {journal} {\bibinfo  {journal} {Annual review of materials
  research}\ }\textbf {\bibinfo {volume} {32}},\ \bibinfo {pages} {113}
  (\bibinfo {year} {2002})}\BibitemShut {NoStop}%
\bibitem [{\citenamefont {Cahn}(1961)}]{Cahn61On}%
  \BibitemOpen
  \bibfield  {author} {\bibinfo {author} {\bibfnamefont {J.~W.}\ \bibnamefont
  {Cahn}},\ }\href {\doibase https://doi.org/10.1016/0001-6160(61)90182-1}
  {\bibfield  {journal} {\bibinfo  {journal} {Acta Metallurgica}\ }\textbf
  {\bibinfo {volume} {9}},\ \bibinfo {pages} {795 } (\bibinfo {year}
  {1961})}\BibitemShut {NoStop}%
\bibitem [{Sup()}]{Sup}%
  \BibitemOpen
  \href@noop {} {}\bibinfo {note} {See Supplemental Material for a detailed
  description of the Landau free energy of $\mathrm{VO}_2$ and a detailed
  derivation of the photoexcitation rate of free electron-hole
  pairs.}\BibitemShut {Stop}%
\end{thebibliography}%

\end{document}